\documentclass[aps,pre,twocolumn,showpacs,amssymb]{revtex4}

\usepackage[dvips]{graphicx}
\usepackage{dcolumn}
\usepackage{bm}
\def\be{\begin{equation}}
\def\en{\end{equation}}

\def\p{\partial} 
\newcommand{\av}[1]{\langle{#1}\rangle}
\def\gs{\gtrsim}
\def\ls{\lesssim}
\newcommand{\bi}[1]{\mbox{\boldmath$#1$}}
\def\p{\partial}
\def\bea{\begin{eqnarray}}
\def\ena{\end{eqnarray}}
\newcommand{\ppp}[3]{{\bigg(}\frac{\partial {#1}}{\partial {#2}}{\bigg )}_{#3}}

\begin{document}


\title{Molecular Dynamics Simulation of 
Heat Conduction in Near-Critical Fluids}


\author{Toshiyuki Hamanaka}
\affiliation{Department of Physics, Kyoto University, Kyoto 606-8502,
Japan}

\author{Ryoichi Yamamoto}
\affiliation{Department of Physics, Kyoto University, Kyoto 606-8502,
Japan}

\author{Akira Onuki}
\affiliation{Department of Physics, Kyoto University, Kyoto 606-8502,
Japan}


\date{\today}

\begin{abstract}
Using molecular dynamics simulations  
we study supercritical 
fluids near the gas-liquid critical point 
under heat flow in two dimensions.  
We calculate the steady-state temperature and 
density profiles. The resultant thermal conductivity 
exhibits critical singularity 
in agreement with the mode-coupling theory in two 
dimensions.  We also calculate  
distributions of the momentum   and 
heat fluxes at fixed density. They 
indicate   that  liquid-like (entropy-poor)  
clusters move toward the warmer boundary and  
gas-like (entropy-rich) 
regions move toward the cooler boundary in a temperature 
gradient.  This counterflow  results  in  critical enhancement of 
the thermal conductivity.

\end{abstract}


\maketitle


\section{INTRODUCTION}

In one-component fluids, 
the density and energy 
 fluctuations are known to become long-ranged 
and long-lived   as the temperature $T$ and the density $n$ 
 approach  the critical values  $T_{\rm c}$ and 
$n_{\rm c}$ \cite{Sengers,onuki}. The critical singularities 
are characterized by the  correlation length 
$\xi$,  which grows as  $\xi_0 (T/T_{\rm c}-1)^{-\nu}$ 
on the critical isochore with $\xi_0$ being 
a microscopic length and 
$\nu$ being the critical exponent. 
On one hand, 
the isothermal compressibility $K_T$, the isobaric 
thermal expansion coefficient $\alpha_p$,  
and the isobaric specific heat $C_p$ 
grow as $\xi^{2-\hat{\eta}}$,  
with $\hat{\eta}$ being the small Fisher critical exponent. 
On the other hand,   the thermal 
diffusivity $D_T$ behaves  as $k_{\rm B}T/\eta \xi^{d-2}$ 
and the life time of the critical fluctuations 
grows as $\tau_\xi= \xi^2/D_T \sim \xi^d$, where   
$d$ is the space dimensionality and the 
weak singularity of the shear viscosity 
$\eta$   is neglected. As a result, 
 the  thermal  conductivity  
$\lambda=D_TC_p$ grows as $\xi^{4-d -\hat{\eta}}$.
The  critical behavior of $\lambda$ and $\eta$ has 
  been well described  by  
the mode-coupling theory \cite{Kadanoff,Kawasaki} and 
by  the dynamic renormalization group theory \cite{Siggia}.

However, the calculations in these dynamical  theories 
are performed in the 
space of the wave vector of the fluctuations 
and are rather formal. 
The real space picture 
of the enhanced heat transport
   in a small temperature 
gradient $dT/dz$ is as follows \cite{onuki}. 
The critical fluctuations with 
relatively higher (lower) 
densities should be  convected in the direction 
(reverse direction) of the temperature gradient. 
The  typical velocity of the clusters with lengths of order $\xi$  
is given by 
\be 
v_\xi \sim   
(\xi/\xi_0)^{\beta/\nu+2-\hat{\eta}}\frac{D_T}{T} \frac{dT}{dz},   
\en  
in the linear response with $\beta$ being the critical exponent.  
The  entropy of the 
liquid-like  regions is smaller  than that 
of the gas-like regions by  $n\xi^d \delta s 
\sim \xi^{d-\beta/\nu}$, where 
the entropy fluctuation 
$\delta s$ (as well as the denisity 
fluctuation $\delta n$) has sizes typically  of order 
$\xi^{-\beta/\nu}$ ($\sim (T/T_{\rm c}-1)^\beta$ on the critical 
isochore). The thermal average of the 
convective heat flux $nT\delta s v_\xi$ 
thus gives rise to   the critical  
heat conduction. 
Within their life times the  clusters 
 can move only over the distance,   
\be 
v_\xi \tau_\xi
 \sim  \frac{\Delta n}{n} (\xi/\xi_0)^{\beta/\nu} 
{\xi^2}/{L}  , 
\en 
where $\Delta n = \alpha_p LdT/dz$  
is the difference  between the densities at the two ends 
of the cell. This distance is very short 
for    $L$ much longer than $\xi$.  
Hence it should be difficult 
to unambiguously  observe the 
 the  cluster motion in heat flow 
experimentally and even numerically.

As numerical work of heat conduction in fluids  \cite{hansen,allen}, 
the thermal conductivity 
has been  calculated  using equilibrium  
MD simulations \cite{a,b} on the basis of the 
Green-Kubo formula \cite{onuki,hansen}  or 
using nonequilibrium MD simulations \cite{d,e,f}. 
In particular, developing   a simple method,  
  Ohara  performed 
nonequilibrium MD simulations for Lenard-Jones (LJ) fluids \cite{ohara} 
and for liquid water \cite{oharaw}. All these previous 
papers treated fluids far from the critical point. 
In this paper we will use 
Ohara's  method to realize heat-conducting states 
in the one-phase region near the critical point.

This paper is organized as follows. 
In Section 2 we will present numerical results 
on  equilibrium critical behavior  in supercritical 
 LJ fluids. 
In Section 3 we will 
 show numerical results on near-critical 
heat conduction together 
 with theoretical interpretations. 
In particular, 
we will confirm the cluster convection 
mechanism  by  introducing steady-state 
distributions of 
the momentum  and  heat fluxes  
at fixed density. 
In Appendix B  we will summarize 
  the mode-coupling theory 
for the thermal conductivity. In Appendix C  
we will examine 
the linear response to  heat flow 
and  justify Eq.1.

\section{model and equilibrium results}

We used  a  two-dimensional (2D) LJ fluid 
composed of $N$ identical particles. 
The pair potential as a function of 
 the distance $r$ between two  particles  
is given by 
\begin{equation}
\phi (r)=4\epsilon \left[ \left(\frac{\sigma}{r}\right)^{12}
-\left(\frac{\sigma}{r}\right)^{6}\right] -C \quad (r\le r_{\rm c}),
\end{equation}
and $\phi(r)=0$ for  $r > r_{\rm c}$. 
The constant $C$ is chosen such that $\phi(r_{\rm c})=0$. 
The cut-off length $r_{\rm c}$ was set equal to $3\sigma$.  
The system contains $N=5000$ particles. 
Space and time are measured in units of $\sigma$ and 
$\tau_0=({m\sigma^{2}/\epsilon})^{1/2}$, where $m$ is the particle mass. 
Equilibrium  states  of the fluid 
may  be characterized by the 
temperature $T$ and the average number density $n= N/V$ 
measured in units of $\epsilon/k_{\rm B}$ and 
$\sigma^{-2}$, respectively. 
 The pressure is measured in units of 
$\sigma^{-2}\epsilon$. 
We used the leapfrog algorithm 
 to integrate the Newton  differential
equations with a time step of $0.01$ 
and the  cell-index method   
of  cutting off  the interaction 
potential. The details of these 
numerical methods  are 
described in the  literature \cite{allen}.

The phase diagram of the 
two-dimensional LJ fluid has been studied 
by several groups using  the 
conventional Monte Carlo method \cite{barker}, 
the Gibbs ensemble method 
\cite{singh,smit,jiang}, and finite-size 
scaling analysis \cite{rovere1993}.
Table  \ref{cp}  summarizes 
the critical parameters 
reported in the literature 
together with our MD results. 
However, note that the critical parameters 
largely depend on the details of the truncation of the 
potential \cite{smit}. 
We also mention 
that Luo {\it et al} \cite{luo} 
examined thermal relaxation 
in  a two-dimensional 
supercritical LJ fluid.

\begin{table}
\caption{
Numerical estimations for  
the two-dimensional LJ fluids.
The estimated critical 
temperature $T_{\rm c}$ and density $n_{\rm c}$ 
are given 
in the first and second columns. 
The particle number 
 $N$ and the cut-off radius 
$r_{\rm cut}$ used are given in 
 the third and fourth columns.
}
\label{tab:cp}
\begin{ruledtabular}
\begin{tabular}{dddcc}
\mbox{$T_{\rm c}$} & \mbox{$n_{\rm c}$} & \mbox{$N$} & \mbox{$r_{\rm cut}$} & Source \\
\hline
0.533 & 0.335 &  256      &       & \cite{barker} \\
0.472  & 0.33  & 500 &  $\infty$\tablenotemark[1] & \cite{singh} \\
0.515  & 0.355  & 512 &  $\infty$\tablenotemark[1]   & \cite{smit} \\
0.459  & 0.35  & 512 &  2.5 & \cite{smit} \\
0.498  & 0.36  & 8000 &  $\infty$\tablenotemark[1]   & \cite{jiang} \\
0.47  & 0.35  &  4096 & 2.5 & \cite{rovere1993}\\
0.47   & 0.37  & 5000 & 3.0 & Present work \\
\end{tabular}
\end{ruledtabular}
\tablenotetext[1]{The Gibbs ensemble method was used
with $r_{\rm cut}=L/2$, where $L$ is the cell size.
}
\label{cp}
\end{table}

As preliminary work before nonequilibrium 
simulations, 
 we carried out equilibrium 
 simulations in the canonical (constant-$NVT$)
ensemble,  using the Nose-Hoover thermostat \cite{allen,nose} 
under  the periodic boundary condition, with 
 $\Delta t=0.01$,  
 to calculate 
the structure factor $S(q)$.   
We started  with  
random initial particle configurations 
at each  given temperature,  
waited for $t_{\rm w}=5\times  10^4$, 
and afterward took data  in 
a subsequent time interval of 
$t_{\rm w}<t< 2t_{\rm w}$.  
This long equilibration is needed 
 because  the density fluctuations 
relax  very slowly  
near the  critical point. That is,  
$t_{\rm w}$ should be longer than 
the life time $\tau_\xi = \xi^2/D_{\rm T}$ 
of   the critical fluctuations with sizes of 
the order of the correlation length 
$\xi$ \cite{onuki}.  Here 
 $\lambda$ is the thermal conductivity and 
 $C_p$ is the isobaric heat capacity per unit volume, 
respectively. In particular, in 
two dimensions the critical exponent $\nu$ is equal to 1 
and the critical singularity of  
$D_T$ is weak (as will be evident in Eq.19), so   
on the critical isochore $\tau_\xi$ grows as 
\be 
\tau_\xi \sim \xi^2 \sim (T/T_{\rm c}-1)^{-2}    \quad ({\rm d}=2).
\en

We consider the structure factor  given by 
\be 
S(q)=\int d{\bi r}
e^{i\small{{\bi q}}\cdot{\small{\bi r}}}
\av{\hat{n}({\bi r},t)\hat{n}({\bi 0},t)}/n. 
\en
Here we define the fluctuating particle number density 
in terms the particle positions as  
\be
\hat{n}({\bi r},t)=\sum_{i=1}^{N}\delta({\bi r}_{i}(t)-{\bi r}) . 
\en
We took  the angle average in the calculation of $S(q)$. 
Fig.1  shows $S(q)$  for  $T=0.65$, 
$0.51$, $0.5$, $0.495$, and $0.49$ 
at $n=0.37$.
We  can see the power law  $q$ behavior,
\be 
S(q)  \sim q^{-7/4},
\en 
in the range $\xi^{-1} \ls q \ls 2$ 
near the critical point, where the exponent value $7/4$ 
is consistent   with the well-known 
Fisher critical exponent $\hat{\eta}=1/4$  
in two dimensions. The peak around $q \sim 6$ represents 
the short-range  pair correlation at this density.  
We then determined the correlation length 
$\xi$ by fitting the data 
to the extrapolated expression  
$S(q)=n k_{\rm B}T K_T /[1+(q\xi)^2]^{7/8}$ for $q \ll  1$. 
The isothermal compressibility 
$K_T= (\p n/\p p)_T/n $ can be determined 
from the long wavelength limit of $S(q)$. 
We show $\xi$ vs  $n$ in Fig.2 and 
$K_T$ vs  $n$ in Fig. 3 
 for  various $T (\ge 0.49)$.   Although not shown 
in Figs.2 and 3, we also performed simulations at lower 
temperatures to  obtain  
$\xi \sim 50$ for $T=0.485$ and 
$\xi >L$  (apparently) for $T=0.48$  at $n=0.37$. 
However, for these $T$,  
 our simulation times are not sufficiently long 
compared with $\tau_\xi$ in Eq.4. 
From the peak positions   
in Figs.2 and 3 
we estimated $n_{\rm c} \cong  0.37$ 
in Table 1.  This value was also obtained 
as the mean position of the two peaks in 
the one-body density distribution $\Psi(\rho)$ (defined by Eq.22 below)
in equilibrium for $0.495\le T \le 0.50$.

From the data in Figs.2 and 3  on the critical isochore,   
 $K_T$ behaves as a function of $\xi$ as 
\be 
K_T= 3.7 \xi^{7/4}+0.80,
\en 
in units of $\sigma^2/\epsilon$. 
We then fitted 
$\xi$ to  the scaling form 
$\xi= \xi_0(T/T_{\rm c}-1)^{-1}$ on the critical isochore to 
obtain $T_{\rm c}=0.47$ and $\xi_0=0.6$. 
From the isothermal curves in the $p-n$  plane in the range  
$0.495 \le T\le 0.50$ at $n=n_{\rm c}$, 
we obtained\cite{comment}      
\be 
\ppp{ p}{ T}{n}  \cong 
\ppp{ p}{T}{\rm cx} \cong 0.40.
\en  
We also consider  the  specific heat 
$C_p= n T (\p s/\p T)_p$ per unit volume ($s$ being the entropy 
per particle) and the thermal 
expansion coefficient $\alpha_p= - 
(\p n/\p T)_p/n$ at constant pressure.  These quantities 
grow strongly and  are related to 
$K_T$ by \cite{onuki}  
\be 
C_p\cong  T\ppp{ p}{ T}{\rm cx}^2K_T, 
\quad    
\alpha_p\cong \ppp{ p}{ T}{\rm cx}K_T,  
\en 
near the critical point. These relations will be used 
in the next section.

As long as $\xi \ll L$,  
our equilibrium results are consistent  with the 
well-known results of critical phenomena 
\cite{Sengers,onuki}. If $\xi$ approaches $L$, 
the finite-size scaling analysis may be performed  
\cite{rovere1993}. However, such analysis is beyond the scope 
of this paper.

\section{Nonequilibrium simulations}

\subsection{Method} 

Next we imposed  a heat flux to the system 
using Ohara's method \cite{ohara,oharaw}. 
As illustrated  in Fig.4, the cell 
is divided into three parts, cooling, heating, and interior 
regions. 
In the cooling region $-0.5L<z<-0.4L$ 
the average temperature  of the particles  
was  kept at $T_{\rm L}$, while 
 in the heating region 
$0.4L<z<0.5L$  it was  kept at $T_{\rm H}$. 
The precise definition of 
the average temperature in a given  region 
will be presented  in Eq.12 below.  The 
 pinning of the average temperatures  in the cooling and heating 
regions was realized by simple scaling of the velocities 
of the particles in the two 
regions at every time step.   
The periodic boundary condition 
was imposed in the $x$  direction, 
while the walls  at $z= \pm L/2$  
were assumed to interact with 
the particles via the LJ 
potential in Eq.3 where  $r$ is the distance 
from the wall  and  $r_{\rm c}=3$. The particles in the interior 
 ($-0.4L < z < 0.4L$) obeyed the Newtonian dynamics without 
artificial thermostat.  The particles entering the interior 
from the cooling (heating) region 
have lower (higher) kinetic energies than  those of the 
particles in the interior on the average. Then a steady heat conducting state 
is realized  after a transient time.

In our  nonequilibrium simulations, 
we used 
a single density $n=0.37$ nearly equal to 
$n_{\rm c}$.  The system length is then 
 $L=(5000/0.37)^{1/2}= 116$. 
The lower boundary 
temperature $T_{\rm L}$ was changed  as 
$0.7$, $0.65$, $0.6$, $0.52$, $0.505$, $0.5$, $0.495$, 
and $0.49$.
The  temperature difference 
$\Delta T=T_{\rm H}-T_{\rm L}$ was fixed at $0.005$ 
in all the simulations. 
We regard  the system 
to be  in a steady state for 
 $t> t_{\rm w}= 6\times 10^4$ 
after  application of $\Delta T$.  In the following 
the steady-state values of the physical quantities 
will be the  time averages  over the  
data during the next time interval 
$t_{\rm w}<t<t_{\rm w} +t_{\rm data}$  
with $t_{\rm data}=14 \times 10^4$.

\subsection{Steady-state  density and temperature 
profiles}

Fig.5 displays  a snapshot of the particle positions 
in the cell at $t=2\times 10^5$  
for $T_{\rm L}=0.50$, where  the system is nearly 
in a steady state.   
The large clusters formed by many particles 
 are significantly denser 
near  the cooler boundary (bottom) 
than near the warmer boundary (top)\cite{commentd}. 
This is due to the diverging 
isobaric thermal expansion as 
will be shown in Eq.16 below. 
By comparing successive snapshots (not shown here), 
we recognize that the clusters  
appear and disappear continuously 
on the time scale of $\tau_\xi$ 
in Eq.4.

 To quantitatively analyze Fig.5, 
we  need to calculate  the time averages of the  
temperature and the density. 
They are defined as follows. 
Dividing  the interior 
into eight layers 
with thickness $L/10$, 
the density in the $\ell$-th layer is 
defined by 
$n_\ell (t) =({10}/{L^2})N_\ell(t)$ in terms of the 
particle number in the $\ell$-th layer,   
\be 
N_\ell (t) =  
\int_{z_\ell}^{z_{\ell+1}}dz 
\int_0^L   dx
\hat{n}({\bi r},t) ,
\en 
where 
$z_\ell= (\ell-5) L/10$ and 
 $\hat{n}({\bi r},t)$ 
is the fluctuating density in Eq.6.
The temperature in the $\ell$-th layer 
$T_\ell(t)$ may be  defined by 
\be 
T_\ell (t) = \frac{1}{N_\ell(t)} 
 \sum_{i \in \ell}
 |{\bi v}_i(t) - \bar{\bi v}_\ell (t)|^2 ,
\en 
where the summation is over the particles within the 
$\ell$-th layer and 
$
\bar{\bi v}_\ell (t) = 
 \sum_{i \in \ell}
{\bi v}_i(t) /{N_\ell(t)}$ 
is the average velocity within 
the $\ell$-th layer. 
Notice that the 0-th layer 
is  the cooling region 
 and the 9-th layer is the 
heating region. Thus 
we set  $T_0(t)=T_{\rm L}$ 
and $T_9(t)=T_{\rm H}$  in the cooling and heating regions,  
respectively.

Fig.6 shows  the steady-state  temperature and 
density profiles, 
\be 
T(z)= \av{T_\ell(t)}, \quad 
n(z)= \av{n_\ell(t)},
\en  
which are the time averages 
 with $z = (\ell-5) L/10$. 
Here $T_{\rm L}=0.50$ and 
$n=0.37$  as in Fig.5. 
For this case the deviation of 
$T(z)$   from the linear profile is not large 
and there exists a temperature gradient also 
in the cooling and heating regions. 
We may define the penetration widths 
$d_{\rm L}$ and $d_{\rm H}$ by  the extrapolations, 
\be    
T(-d_{\rm L}-2L/5)=T_{\rm L}, \quad 
T(2L/5+d_{\rm H})=T_{\rm H}.
\en  
When the temperature profile is nearly linear, 
we simply find 
 $d_{\rm L} \cong d_{\rm H} \cong L/20$.  
When the  temperature flux  is not too large, 
 the effective cell length  of heat conduction becomes 
\be 
4L/5+d_{\rm L}+d_{\rm H} \cong 9L/10 , 
\en 
which is the distance   between 
the middle points of 
the cooling and heating region 
as in the case of  
Ohara's simulation \cite{ohara}.

On the other hand, in Fig.6 
the density deviation is much more 
enhanced than that  of the temperature. 
We expect that 
if the deviation $\delta T(z)= 
T(z)-T(0)$ (measured from the center $y=0$)  
is not too large, 
the average density deviation 
$\delta n(z)= 
n(z)-n(0)$ 
should be  given  by 
\be 
\delta n(z)  \cong - n\alpha_p {\delta T(z)}, 
\en 
where $\alpha_p$ is the isobaric thermal 
expansion coefficient. Here,  at the center, 
we find $T(0)=0.5025$ and $n(0) =0.37$, so 
 Eq.10 indicates 
$K_T= 159$ and $n\alpha_p= 23.5$ at the center. 
In Fig.6 the  solid line 
 has a slope of $\Delta T/(9L/10)= 0.0056/L$, 
while the dotted line 
a  slope  of $-23.5 \times 0.0056/L$. 
These two lines can fairly fit 
the temperature and density data, though  
there are considerable deviations close to 
the boundaries (obviously because 
the boundary regions are considerably off-critical). 
 Notice that 
we assume homogeneity of the pressure in Eq.16. 
To check this, we calculated 
 the steady-state time averages  
of the trace of the stress tensor 
(integrated in each layer)    
and found no appreciable heterogeneity 
 in these    values.

\subsection{Thermal conductivity}

In our simulations, 
the steady-state  thermal conductivity   $\lambda$ 
in the interior was  calculated from  
\begin{equation}
\lambda=Q \frac{0.9L}{\Delta T}, 
\end{equation}
where $0.9L$ is the effective cell length in Eq.15.
The  $Q$ is the steady-state heat flux written as  
\be 
Q= - {\av{J^{Qz}_0(t)}}/ {0.8L^2}
\en 
where  $J^{Qz}_0(t)$ is   the integral of the heat flux 
density within the interior with area $0.8L^2$ 
and its microscopic expression 
will be given in Appendix A.    
In our small system 
the fluctuations of the heat flux turned out to be 
large, so we performed 10 independent runs 
and calculated  the mean values 
of the corresponding 10 time averages.  
In Fig.7   
 the thermal conductivity data are shown 
as a  function of the temperature at the critical density, 
which  gives the background 
$\lambda_{\rm B}= 2.3$ far above the critical point.
In Fig.8   the data of the 
singular part 
$\Delta \lambda=\lambda-\lambda_{\rm B}$ are plotted 
 as a functions of $\xi$.   For $\xi \ls 10$ 
our numerical data nicely agree with the 
theoretical linear response result Eq.19, 
which will be explained below.
For $\xi \gs 10$ the finite-size effect and 
the nonlinear response effect should 
be responsible for the saturation of the 
calculated $\lambda$.

The mode-coupling theory 
in Appendix B predicts the following behavior,
\bea 
\lambda &=& 
\lambda_{\rm B}+ \frac{T}{4\pi\eta}C_p \ln (L/\xi) \nonumber\\
&=& \lambda_{\rm B}+ A_\lambda \xi^{7/4} \ln (L/\xi) , 
\ena 
where  $\eta$ is the shear viscosity \cite{tail}. 
See also Appendix C for the linear response theory for heat flow. 
The singular part of the thermal conductivity is simply given by 
 $D_T C_p$ with $D_T$ being 
the thermal diffusion constant \cite{Kawasaki}. 
In terms of 
the isothermal compressibility 
 $K_T$ the coefficient  $A_\lambda$ is written as 
\be 
A_\lambda= \frac{T^2}{4\pi\eta} 
\bigg ( \frac{\p p}{\p T} \bigg )_{\rm cx}^2 
K_T \xi^{-7/4} . 
\en   
To estimate $A_\lambda$ from the above expression,  
we  calculated the shear viscosity $\eta$ 
at $T=0.50$ and $n=0.37$ by two methods 
and obtained almost the same results. 
That is, (i) the time integral of the stress 
time-correlation function 
\cite{hansen,allen} in the range $0<t<100$ gave   
$\eta \cong 0.35$  and  
(ii) the long time tail of the velocity 
correlation function  gave $\eta \cong  0.33$ (see Appendix D). 
If we use  the latter  result together with Eqs.7 and 8, 
we are led to $A_\lambda =0.035$. 
In Fig.8 the theoretical curve represents 
the second term on the right hand side of 
Eq.19 with this $A_\lambda$. It excellently 
agrees with the data before the saturation of $\lambda$.

\subsection{Momentum and heat flux distributions 
at fixed density under heat flow}

Some characteristic features of the density  fluctuations 
can be seen in 
 the one-body 
density distribution function 
$\Psi (\rho) =\av{\delta( {\hat{n}}_{\rm cell}-\rho)}$,  
where $\hat{n}_{\rm cell}$ is the density in an 
appropriately chosen cell in the fluid 
and the average is taken over the thermal fluctuations 
and  over the cells in the system  
\cite{rovere1993,smit,jiang,Wilding}. 
It is the probability distribution 
of the coarse-grained density.  Furthermore, 
in the presence of heat flow, 
we are interested in  distributions of 
the momentum  and  heat fluxes  at fixed 
density. They  can give  the correlations 
between these fluxes 
and the  density within the same cell 
even if the cluster motion driven by heat flow 
is very  small.

First we coarse-grain the system 
to calculate  $\Psi (\rho)$.  
The interior region 
 ($-0.4L < z < 0.4L$ and $0<y<L$)  is divided into 
$10\times10$ rectangular subsystems. 
Let  $M_k(t)$ ($k=1, \cdots, 100$) 
be the  particle number in the 
$k$-th cell at time $t$. 
After the time averaging  in steady states,   
we obtained 
the distribution of $M_k(t)$ for integer $M$ as 
\be 
P(M) = \frac{1}{100}  \sum_{k=1}^{100}
\av{ \delta_{M, M_k(t)} } , 
\en 
where $\delta_{M,M'}$ is the Kronecker delta, 
and $\sum_{M=0}^{\infty}P(M)=1$  by definition. 
For each  given density $\rho= M/V_{\rm cell}$ 
we define 
\bea 
\Psi(\rho) &=&
 V_{\rm cell} P(V_{\rm cell}\rho)\nonumber\\
&=& \frac{1}{100}  \sum_{k=1}^{100}\av{
 \delta(\rho -  n_k(t))}, 
\ena 
where $V_{\rm cell}= 0.8 L^2/100$ is the cell volume 
and the second line is the expression in the 
continuum limit with  
$n_k(t)= M_k(t)/V_{\rm cell}$. By definition 
we obtain 
\be 
\int_0^\infty  d\rho \Psi(\rho)=1, 
\quad \int_0^\infty  d\rho \rho \Psi(\rho)=
n_{\rm in},
\en 
where $n_{\rm in}$ is the 
average density in the interior and 
 $n_{\rm in}\cong n$ in our case. 
The second moment 
becomes 
\be 
\int_0^\infty d\rho (\rho-n_{\rm in})^2 \Psi(\rho)=
\frac{1}{100}  \sum_{k=1}^{100}\av{
 (n_k(t)-n_{\rm in})^2 }.  
\en 
In equilibrium, or 
if the heterogeneity along the heat flow 
is neglected, the second moment 
behaves as $\xi^{2-\hat{\eta}}/V_{\rm cell}$ 
for $\xi $ less than the cell length but 
 as $V_{\rm cell}^{(2-d -\hat{\eta})/d}$ 
for larger $\xi$ due to the finite-size effect.

Now we consider   the coarse-grained 
momentum and heat fluxes 
at fixed density. We calculate the 
following  steady-state averages, 
\begin{equation}
J_p(\rho)=
\frac{1}{100V_{\rm cell}}
\sum_{k=1}^{100}
\langle J^{z}_{0k}(t)\delta(\rho - n_k(t))\rangle
\end{equation}
\begin{equation}
J_Q(\rho)=
\frac{1}{100V_{\rm cell}}
\sum_{k=1}^{100}
\langle J^{Qz}_{0k} (t)\delta(\rho-n_k(t))\rangle,
\label{ddist}
\end{equation}
where $J^{z}_{0k}(t)$ and 
$J^{Qz}_{0k} (t)$ are the $z$ component 
of the space integral of 
the momentum density and that of 
the heat flux, respectively,  
within the $k$-th cell (see (A.2) in the Appendix A 
for their definitions). 
If they are divided by the cell volume 
$V_{\rm cell}$, they 
become the cores-grained 
 densities, respectively. For simplicity, 
we  may write $\Psi(\rho)= \av{\delta(\rho-\hat{n})}$,  
 $J_p(\rho) = \av{J_z\delta(\rho-\hat{n})}$ 
and $J_Q(\rho) = \av{J_z^Q\delta(\rho-\hat{n})}$ 
regarding the dynamic variables involved as the coarse-grained 
quantities.  
The normalized quantities 
$J_p(\rho)/\Psi(\rho)$ and 
$J_Q(\rho)/\Psi(\rho)$ may be interpreted as 
 the coarse-grained conditional 
average of the momentum density and that of 
the heat flux, respectively, 
under the condition of fixed  
density at $\rho$.  
If integrated over $\rho$, we obtain  
\be 
\int_0^\infty  d\rho J_p(\rho)=0 ,
\en 
\be 
\int_0^\infty  d\rho J_Q(\rho)= - Q, 
\en 
where $Q$ is the average heat flux defined by 
Eq.18  in the interior.
In Appendix C we will examine  
the expected behavior of 
these quantities using the 
 linear response theory for $\nabla T$ 
\cite{Oppenheim}.

In Fig.9 we show the three quantities, 
$\Psi(\rho)$,  $J_{p}(\rho)$, 
 and $J_{Q}(\rho)$,  
obtained from 10 independent runs. 
The temperature at $z=L$ is 
 $T_{\rm L}=0.65$ 
in (a) (upper plate), 
 $T_{\rm L}=0.5$ in (b) (middle plate), and  
$T_{\rm L}=0.48$ in (c)(lower plate),  
with $\Delta T=0.005$ or $\av{dT/dz}=0.43\times 10^{-4}$.  
As can be seen in Fig.7,  
the calculated thermal conductivity 
is $\lambda=5.96$   in (a),  $5.66$ in (b), and $2.63$ in (c). 
 Salient features are as follows.\\ 
(i) The density distribution $\Psi(\rho)$ 
has a rather sharp peak in (a), 
a broad (still single) peak in (b), and 
double peaks in (c).  We also calculated 
$\Psi(\rho)$ in equilibrium at the same temperatures, 
which exhibits   double flattened peaks 
for  $T=0.5$  and sharper double 
peaks for $T=0.48$, so  
 the double peak behavior  emerges  more 
conspicuously  in equilibrium. 
Furthermore, as a complicating factor in heat flow, 
Fig.5  indicates that  
the  average density profile 
is considerably dependent on $z$  
in (b) and (c).\\ 
(ii) The momentum distribution $J_{p}(\rho)$ 
is positive for $\rho \gs 
0.37$ and  negative for for $\rho \ls 
0.37$. This is consistent with 
 the anti-symmetric  behavior, 
$
J_{p}(\rho) \sim    Q (\rho-0.37) \Psi(\rho),  
$  close to the criticality in  Eq.(C.5)  of Appendix C.  
Evidently,    the  liquid-like  clusters 
move toward the higher temperature boundary, 
while the particles in the gas-like regions  
move  toward the lower temperature boundary.  
However, notice that the high-density maximum is  considerably 
sharper than the low-density minimum, which should arise from 
the gas-liquid asymmetry of the fluctuations \cite{comment}. 
In particular, for the case (b), 
the momentum density  of the liquid-like regions 
is of order $10^{-3}$ and   the velocity is of order 
$3\times 10^{-3}$ (in units of $\sigma/\tau_0= 
(\epsilon/m)^{1/2}$). In this case we have 
$\xi \sim 18$ and $D_T \sim 0.1$ 
so that  the distance of the cluster motion  
within the life time $\xi^2/D_T \sim 3\times 10^3$ 
is estimated to be  of order 10.\\ 
(iii) The heat flux 
distribution function  $J_{Q}(\rho)$ 
 still  exhibits considerable irregular behavior, 
but its negativity at any $\rho$ is clear. 
Let us   smooth out the curves; then,  $J_{Q}(\rho)$ 
 has a  single minimum in (a) and 
double minima in (b) and (c). 
Thus, as $T \rightarrow T_{\rm c}$, 
 heat is largely transported by the  
counterflow of  the  liquid-like  clusters 
and  the gas-like regions. Particularly in (c), 
 the  contribution from 
$\rho \cong 0.37$ becomes  very small and 
 the curve can be fairly fitted to the symmetric 
relation  
$
J_{Q}(\rho) \sim  -Q  (\rho-0.37)^2 \Psi(\rho)$ 
in accord with Eq.(C.6).  
The gas-liquid asymmetry is more suppressed for $J_{Q}(\rho)$ than 
 for  $J_{p}(\rho)$.

\section{concluding remarks}

 MD simulations have been performed on LJ near-critical 
fluids  in two dimensions. In equilibrium 
the critical properties obtained are presented in Figs.1-3. 
The main results under heat flow   are 
summarized as follows.\\
(i) We have calculated the 
 average density and temperature profiles in a steady state 
in Fig.4, where they are fairly fitted to  linear 
lines and satisfy Eq.16. The density deviation is much enhanced 
than that of the temperature and the average pressure remains 
homogeneous.\\
(ii)  We have  obtained    critical  enhancement 
of the thermal conductivity 
for various $T$ close to $T_{\rm c}$ in Figs.7 and 8 
in good agreement with 
the mode-coupling  prediction in Eq.19 derived 
in Appendix B.\\
(iii) We have calculated the 
one-body density distribution $\Psi(\rho)$, 
the momentum distribution  $J_p(\rho)$,  and 
the heat flux distribution $J_Q(\rho)$ defined 
by Eqs.21, 24, and 25. 
Fig.9 demonstrates  the cluster convection mechanism,  
which is briefly summarized in the introduction and 
supported in Appendix C in the linear regime.\\ 
(iv) The  cluster  convection is  a natural consequence of 
the irreversibility in heat conduction, while 
the  density increase 
near the cooler boundary   in Fig.6 
arises from   the simple thermodynamics under 
homogeneous pressure in Eq.16.  
These two effects are not 
contradictory with each other 
in view of the fact that  the distance of cluster convection 
 is very short.

The following problems  could 
 be mentioned as future subjects 
of  nonequilibrium MD simulations.\\
(i) When the boundary wall is heated with a fixed cell volume, 
 sound waves emitted from the boundary 
 can cause rapid adiabatic 
 heating throughout the cell (the piston effect) 
\cite{Ferrell,Beysens}. We should examine 
how this phenomenon starts  
in the early stage on the acoustic time scale \cite{onuki}.\\
(ii) Heat conduction in two-phase near-critical fluids  
below $T_{\rm c}$ 
has been little examined in the literature \cite{onuki}. 
For example, we should examine  how 
a gas-liquid interface reacts  
to applied heat flow, where latent heat transport 
can be crucial in the presence of convection.
Interestingly,  gas bubbles in liquid migrate  
toward the warmer boundary in heat flow 
owing to the Marangoni effect  
\cite{Beysens1}.@

\begin{acknowledgments}
This work is supported by 
Grants in Aid for Scientific 
Research 
and for the 21st Century COE project 
(Center for Diversity and Universality in Physics)
 from the Ministry of Education, 
Culture, Sports, Science and Technology of Japan.
Calculations have been performed at the Human Genome Center, 
Institute of Medical Science, University of Tokyo and 
the Supercomputer Center, Institute for Solid State Physics, 
University of Tokyo. 
\end{acknowledgments}

\vspace{2mm} 
{\bf Appendix A: Microscopic Expressions}\\
\setcounter{equation}{0}
\renewcommand{\theequation}{A.\arabic{equation}}

We introduce  the momentum 
density, 
\be 
{\bi J}({\bi r},t)= 
\sum_i m{\bi v}_i(t) 
\delta({\bi r}-{\bi r}_i(t))
\en  
and the energy current  density 
${\bi J}^{\rm e}({\bi r},t)$. 
The microscopic expression for the latter quantity 
is rather complicated \cite{onuki}. 
Let us consider its 
 space integral 
$ {{\bi J}}^{\rm e}_0(t)
= \int_{V_1} d{\bi r} {{\bi J}}^{\rm e}({\bi r},t)$ 
in  a subsystem with volume $V_1$ 
containing many particles. It  
may be approximated as 
\begin{eqnarray}
{{\bi J}}^{\rm e}_0(t)
&=& \frac{1}{2} {\sum_{i}}' 
 \bigg [ m{v}_i^{2}
  +\sum_{j\neq i}\phi(r_{ij}) \bigg ] 
{\bi v}_i \nonumber\\
&-&\frac{1}{2}{\sum_{i}}' 
  \sum_{j\neq i}\phi'(r_{ij})\frac{1}{r_{ij}}   
({\bi v}_{i}\cdot{\bi r}_{ij}) {\bi r}_{ij}, 
\end{eqnarray}
where ${\bi r}_i={\bi r}_i(t)$ and 
${\bi v}_i={\bi v}_i(t)$ are the position and 
velocity of the $i$-th particle 
(the time $t$ being suppressed in (A.2)), 
${\bi r}_{ij} ={\bi r}_i-{\bi r}_j$, 
$\phi'(r)= d\phi(r)/dr$, and 
the summation ${\sum_i}'$ is over the particles 
contained in the subsystem under consideration. 
Here the pair interactions 
between the particles inside and outside the subsystem 
are not precisely accounted for.

The microscopic heat flux density 
is defined by \cite{onuki}  
\be 
{\bi J}^Q({\bi r},t)= 
{\bi J}^{\rm e}({\bi r},t)
-  [(e+p)/n] {\bi J}({\bi r},t), 
\en 
where $e$, $p$, and $n$ are 
the average energy, pressure,
and density, respectively.
This current satisfies the orthogonal 
property $\int d{\bi r} \av{{\bi J}^Q({\bi r},t)
\cdot {\bi J}({\bi r}',t)}=0$ in equilibrium. 
 The Green-Kubo formula for 
the thermal conductivity reads 
\be
\lambda= \frac{1}{k_{\rm B}T^2}
\int_0^{\infty}dt \int d{\bi r} 
\av{J_{z}^Q({\bi r},t)J_{z}^Q({\bi 0},0)}.  
\en
The ${J}_0^{Qz}(t)$ in Eq.18 is the $z$ component 
of the total heat flux in the interior,  
\be 
{\bi J}_0^{Q}(t) = \int_{\rm interior} 
 d{\bi r}{\bi J}^Q({\bi r},t). 
\en 
In Eqs.24 and 25 the space integrals  
are within small subsystems.

\vspace{2mm} 
{\bf Appendix  B: Mode-Coupling Theory}\\
\setcounter{equation}{0}
\renewcommand{\theequation}{B.\arabic{equation}}

In the critical dynamics of simple fluids 
the gross variables include 
the long wavelength 
parts (with wave numbers in the 
region $q \ll  \sigma^{-1}$)  of 
the  energy density $\hat{e}$, 
 the particle density $\hat{n}$, 
and the momentum density 
$\bi J$.    The heat flux  density ${\bi J}^Q({\bi r},t)$ 
in (A.3) has been approximated as a sum of 
a  product of the gross variables 
and a random part in the form \cite{Kadanoff,Kawasaki,Siggia,onuki},  
\be 
{\bi J}^Q({\bi r},t) = 
\frac{T}{m}  \delta\hat{s} ({\bi r},t) {\bi J}({\bi r},t) 
+{\bi J}_{\rm R}^Q({\bi r},t) .  
\en  
The  $\delta\hat{s} $ is the fluctuating entropy  
deviation (per particle)  defined by 
\be 
\delta\hat{s} ({\bi r},t)= \frac{1}{n T} \bigg [ 
\delta\hat{e} ({\bi r},t)- \frac{e+p}{n } 
\delta\hat{n} ({\bi r},t) \bigg ],  
\en 
in terms of the deviations of the energy density 
$\hat{e}$ and the number density $\hat{n}$ 
The $\hat{e}$ can  be 
defined microscopically using 
the particle positions and velocities 
\cite{onuki,Wilding}.  The first term on the right hand side of (B.1) 
evolves slowly in time and 
 gives rise to the singular part of the thermal conductivity 
$\Delta \lambda$ when substituted into (A.4). In 2D 
the mode-coupling calculation 
yields  the following integral 
over the wave vector $\bi q$, 
\be 
\Delta \lambda= \frac{k_{\rm B}T}{2\eta} 
\int \frac{d{\bi q}}{(2\pi)^2}  \frac{1}{q^2}  C_p(q) , 
\en   
where  $\eta$ is the shear viscosity \cite{tail}  and 
$
C_p(q) =  k_{\rm B}^{-1} 
n^2 \av{|\hat{s}_{\small{\bi q}}|^2} 
$   
is the variance of the entropy fluctuation with   
$\hat{s}_{\small{\bi q}}$ being the Fourier component. 
See Appendix C for another  derivation of $\Delta \lambda$ 
from the linear response. 
As far as the most singular part is concerned, 
we may set \cite{onuki}  
\be 
\delta\hat{s} \cong 
 -n^{-2}(\p p/\p T)_{\rm cx}\delta\hat{n}.  
\en 
This yields 
\be
C_p(q)\cong (\p p/\p T)_{\rm cx}^2 S(q)/k_{\rm B}n,  
\en  
in terms of  the structure factor $S(q)$ in Eq.5 
 \cite{onuki}.
The long wavelength limit 
$C_p= \lim_{q\rightarrow 0}C_p(q)$ is the usual isobaric 
specific heat per unit volume behaving as in Eq.10. 
Note that the integral (B.3) 
is logarithmically divergent 
at small  $q$, so  
we obtain the expression Eq.19. 
On the other hand, the second 
 term  on the right hand side of (B.1) 
relaxes  rapidly and gives rise to the background thermal 
conductivity $\lambda_{\rm B}$.

\vspace{2mm} 
{\bf Appendix  C: Linear Response to Temperature Gradient 
near the Gas-Liquid Critical Point}\\
\setcounter{equation}{0}
\renewcommand{\theequation}{C.\arabic{equation}}

Here we consider 
 the linear response theory 
with respect to a temperature gradient 
${\bi a} \equiv \nabla T$ (along the $z$ axis) 
in a steady heat-conducting state 
in the absence of macroscopic velocity field 
\cite{Oppenheim}. To pick up the singular contribution 
near the gas-liquid critical point 
we may approximate the heat flux  by  
$(T/m)\delta\hat{s} ({\bi r},t) {\bi J}({\bi r},t)$ 
from (B.1). Then the linear response of 
any dynamic variable ${\cal B}({\bi r},t)$ 
to $\bi a$ can  be written as \cite{onuki} 
\bea
\delta \av{{\cal B}}  
&=&   \frac{-{\bi a}}{mk_{\rm B}T}\cdot \int_0^{\infty} 
\hspace{-1mm} dt\hspace{-1mm} 
\int d{\bi r}'
\av{{\cal B}({\bi r},t) 
\delta s({\bi r}',0) 
{\bi J}({\bi r}',0)}\nonumber\\
&&\hspace{-1.1cm}=\frac{\bi a}{mk_{\rm B}T}\cdot \int_0^{\infty} 
\hspace{-1mm} dt\hspace{-1mm} 
\int d{\bi r}'
\av{\tilde{\cal B}({\bi r},0) 
\delta s({\bi r}',t) 
{\bi J}({\bi r}',t)}.  
\label{eq:C.1}
\ena
From the first to second line 
use has been  made of the time-reversal 
 relation  $\av{{\cal A}(t){\cal B}(0)}= 
\av{\tilde{\cal B}(t)\tilde{\cal A}(0)}$ where 
$\tilde{\cal A}$ and $\tilde{\cal B}$ are  
the time-reversed variables. 
For example,   
$\tilde{\bi J}= -{\bi J}$. Furthermore, on the second line, 
we may replace $\delta s({\bi r}',t)$ 
by $\delta s({\bi r}',0)$ 
because the relaxation time  of ${\bi J}({\bi r}',t)$ 
due to the shear viscosity $\eta$ is much faster 
than that of $\delta s({\bi r}',t)$. Then  
the time integral may be performed to give 
\bea 
\delta \av{{\cal B}}&=&   -\sum_{ij} \frac{n a_i}{k_{\rm B}T}  
\int d{\bi r}'\int d{\bi r}''{\cal T}_{ij}({\bi r}'-{\bi r}'')
\nonumber\\  
&&\times \av{{\cal B}({\bi r})
\delta s({\bi r}') 
{J}_j({\bi r}'')},  
 \label{eq:C.2}
\ena
where  the equal-time correlation 
is involved and the time dependence  
is hence suppressed. The 
${\cal T}_{ij} ({\bi r})$
is the Oseen tensor whose Fourier transformation is 
$
{\cal T}_{ij}({\bi q})
 = (\delta_{ij} - q_iq_j/q^2 ) /\eta q^2$. 
In 3D it follows the well-known expression 
${\cal T}_{ij}({\bi r})
 = (\delta_{ij} +x_ix_j/r^2 ) /8\pi\eta r$.

In (C.2), if we set ${\cal B} = (T/m)
\delta\hat{s}  {J}_z$ and use the equilibrium 
relation $\av{J_i({\bi r})J_j({\bi r}')} 
=k_{\rm B}T \rho \delta_{ij}\delta({\bi r}-{\bi r}')$, 
we reproduce  the mode-coupling 
expression for the singular part 
of the thermal conductivity (given by  (B.3) in 2D) 
in the form $\delta\av{\cal B}=-\Delta \lambda dT/dz$. 
Next let us   set ${\cal B} = 
{J}_z\delta(\rho- \hat{n})$ and $
{J}_z^Q \delta(\rho- \hat{n})$ where  
the dynamic variables $J_z$, ${J}_z^Q$, and $\hat n$ 
are  the coarse-grained quantities 
 averaged  in appropriate cells. 
Then  
we obtain $J_p(\rho)$ and $J_Q(\rho)$ in 
Eqs.24 and 25 expressed as  
\bea 
J_p(\rho) &=& m \ppp{p}{T}{\rm cx} \frac{dT}{dz} 
\int d{\bi r}'{\cal T}_{zz}({\bi r}-{\bi r}')\nonumber\\ 
&&\times\av{\delta(\rho-\hat{n}({\bi r})) 
\delta n({\bi r}')}, 
\ena 
\bea 
J_Q(\rho) &=& -\frac{T}{n^2}  \ppp{p}{T}{\rm cx}^2\frac{dT}{dz} 
\int d{\bi r}'{\cal T}_{zz}({\bi r}-{\bi r}')\nonumber\\ 
&&\times\av{\delta(\rho-\hat{n}({\bi r})) 
\delta n({\bi r})  
\delta n({\bi r}')}. 
\ena 
We notice that these quantities   depend on the cell 
volume $V_{\rm cell}$. If the cell length $\ell_{\rm cell}= 
 V_{\rm cell}^{1/d}$ is shorter 
than the correlation length $\xi$, 
we estimate $J_p(\rho)$  as  
\be 
J_p(\rho) \sim  \frac{m}{\eta} \ell_{\rm cell}^{2} 
(\rho-n_{\rm in})    \Psi(\rho) \ppp{p}{T}{\rm cx} 
\frac{\p T}{\p z}     ,  
\en 
where   $n_{\rm in}$ is the average 
density. If $\ell_{\rm cell}$ is longer than $\xi$, 
we divide the cell into subsystems 
with length $\xi$ and  find that 
$J_p(\rho) $ is given by (C.5) with 
$\ell_{\rm cell}^2$ being 
replaced by $\xi^2$.
Next notice that the integral 
$\int d\rho J_Q(\rho) $ is equal to 
$-(\Delta \lambda)dT/dz$ from (C.4) 
which is in accord with Eq.28 for $\Delta\lambda\cong \lambda$.  
Accounting for this sum rule 
 we thus expect  
\be 
J_Q (\rho) \cong -A_Q
(\rho-n_{\rm in})^2    \Psi(\rho) \frac{\p T}{\p z}     ,  
\en 
for $\ell_{\rm cell} \ll \xi$. The coefficient $A_Q$ 
is determined from the normalization condition 
Eq.28. The estimations (C.5) and (C.6) 
are consistent with the data in Fig.9.

In addition,  
Eq.1 in the introduction 
follows  if we assume 
$v_\xi \sim J_p(\rho)/mn\Psi(\rho)$  in (C.5) by setting  
 $\ell_{\rm cell}\sim \xi$ 
and $\rho-n_{\rm in} \sim 
\xi^{-\beta/\nu}$ with the aid of the exponent relation 
$2\beta=(d-2+\hat{\eta})\nu$ \cite{onuki}. 
Note that $J_p(\rho)/\Psi(\rho)$ represents 
the average momentum density at density $\rho$.

\vspace{2mm} 
{\bf Appendix  D: Diffusion in Two Dimensions}\\
\setcounter{equation}{0}
\renewcommand{\theequation}{D.\arabic{equation}}

In two dimensions the flux-time correlation functions for the  
transport coefficients have a long time tail relaxing 
as $1/t$, giving rise to 
a  logarithmic singularity 
(if integrated over time) \cite{tail}.  
The simplest example is the diffusion constant 
$D$ of  a tagged particle. It is the time integral of 
the velocity time-correlation 
function, 
\be 
G(t) = \frac{1}{2N} \sum_{i=1}^N \av{{\bi v}_i(t_0+t)\cdot 
{\bi v}_i(t_0)}.
\en 
The long time tail of $G(t)$ is theoretically 
given by $(k_{\rm B}T/8\pi\eta)/t$ if the kinetic viscosity 
$\eta/mn$ is much larger than $D$. 
By taking the average over  $t_0$ in  
a time interval of $5\times  10^4$, 
we obtained $\int_0^t dt' G(t')\cong 0.17 + 
0.059\log t$ for $t \gs 1$, leading to 
$k_{\rm B}T/8\pi\eta=0.059$. 
Note that the kinetic viscosity is close 1 and 
is considerably larger than the diffusion constant in our system.


\begin{references}


\bibitem{Sengers}
 J.V. Sengers and J.M.H. Levelt Sengers, in {\it Progress in Liquid Physics,}
 edited by C.A. Croxton (Wiley, Chichester, England, 1978). 



\bibitem{onuki}
A. Onuki,
{\it Phase Transition Dynamics} 
(Cambridge University Press, Cambridge, 2002).



\bibitem{Kadanoff} L.P. Kadanoff and J. Swift, 
Phys. Rev. {\bf 166}, 89 (1968).
\bibitem{Kawasaki} 
K. Kawasaki, in {\it Phase Transition and Critical Phenomena}. edited by C. Domb and M.S. Green (Academic, New York, 1976), Vol.5A; Ann. Phys. (N.Y.) {\bf 61}, 1 (1970).
\bibitem{Siggia} E.D. Siggia, B.I. Halperin, and P.C. Hohenberg,
Phys. Rev. B {\bf 13}, 2110 (1976).



\bibitem{hansen}
J.P. Hansen and I.R. McDonald,
{\it Theory of Simple Liquids},
(Academic Press, London, 1986).


\bibitem{allen}  
M.P. Allen and D.J. Tildesley,
{\it Computer Simulation of Liquids} 
(Clarendon Press, Oxford, 1987).






\bibitem{a}  
R. Vogelsang, C. Hoheisel, G. V. Paolini, and G. Ciccotti, Phys. Rev. 
A {\bf 36}, 3964 (1987); 
R. Vogelsang, C. Hoheisel, and G. Ciccotti, J. Chem. Phys. {\bf 86}, 
6371 (1987). 

 
\bibitem{b} 
P. J. Gardner, D. M. Heyes,
 and S. R. Preston, Mol. Phys. {\bf 73}, 141 (1991). 

   

\bibitem{d} 
G. V. Paolini, G. Ciccotti, and C. Massobrio, Phys. Rev. A {\bf 34}, 1355 
(1986).


\bibitem{e}
D. J. Evans and S. Murad, Mol. Phys. {\bf 68}, 1219 (1989);  
B. Y. Wang, P. T. Cummings, and D. J. Evans, Mol. Phys. {\bf 75}, 
1345 (1992). 


\bibitem{f} 
B. Hafskjold, T. Ikeshoji, and S. K. Ratkje, Mol. Phys. {\bf 80}, 1389 (1993); 
T. Ikeshoji and B. Hafskjold. ibid. {\bf 81}, 
251 (1994). 



\bibitem{ohara} T. Ohara, 
J. Chem. Phys. {\bf 111}, 9667 (1999).
\bibitem{oharaw} 
T. Ohara, 
J. Chem. Phys. {\bf 111}, 6492 (1999). 

\bibitem{barker}
J.A. Barker, D. Henderson, and F.F. Abraham,
Physica {\bf 106A}, 226 (1981).



\bibitem{singh}
R.R. Singh, K.S. Pitzer, J.J. de Pablo, and J.M. Prausnitz,
J. Chem. Phys. {\bf 92}, 5463 (1990).

\bibitem{smit}
B. Smit and D. Frenkel,
J. Chem. Phys. {\bf 94}, 5663 (1990).

\bibitem{jiang}
S. Jiang, and K.E. Gubbins,
Molec. Phys. {\bf 86}, 599 (1995).




\bibitem{rovere1993}
M. Rovere,
J. Phys.; Condens. Matt. {\bf 5}, B193 (1993);
M. Rovere, P.Nielaba, and K. Binder,
Z. Phys. {\bf 90}, 215 (1993).

\bibitem{luo}
H. Luo, G. Ciccotti, M. Mareschal, M. Meyer, and B. Zappoli,
Phys. Rev. E {\bf 51}, 2013 (1995).

\bibitem{nose}
S. N\'ose,
Molec. Phys. {\bf 52}, 255 (1983).

\bibitem{comment} 
On the critical isochore 
$(\p p/\p T)_n /(\p p/\p T)_{\rm cx}-1$ is of order 
$C_v/C_p \sim (T/T_{\rm c}-1)^{\gamma-\alpha}$ \cite{onuki}, 
where $\gamma=7/4$ and $\alpha =0$ in two dimensions. 
\bibitem{commentd}
In Ising spin systems 
we may equally define clusters with up-spins 
and those of down-spins. In  our simulations, however,
 only the  denser  clusters ($n>n_{\rm c}$) 
are  well defined  as in Fig.5. 
The symmetry between the  higher-density 
and lower-density critical fluctuations should be attained 
much closer to the criticality.  



\bibitem{Wilding} N.B. Wilding, 
J. Phys. Condens. Matt. {\bf 9}, 585 (1997). 

\bibitem{tail} B. J. Alder and T. E. Wainwright
Phys. Rev. A {\bf 1}, 18 (1970); 
J. R. Dorfman and E. G. D. Cohen, 
Phys. Rev. Lett. {\bf 25}, 1257 (1970);  
T. E. Wainwright, B. J. Alder, and D. M. Gass, 
Phys. Rev. A {\bf 4}, 233 (1971).  
The shear viscosity also 
has a small logarithmic singularity 
$\eta= \eta_{0}+ A_\eta 
\log(L/\sigma)$ with 
$A_\eta/\eta_{0} \sim  mn k_{\rm B}T/8\pi \eta_{0}^2$ 
being  small. 




\bibitem{Oppenheim} I. Procaccia, D. Ronis, 
M.A. Collins, J. Ross,  
and I. Oppenheim, 
Phys. Rev. A {\bf 19}, 1290 (1979).



\bibitem{Ferrell} 
A. Onuki, H. Hao and R. A. Ferrell, Phys. Rev. A {\bf 41}, 
2256 (1990). 

\bibitem{Beysens} 
Y. Garrabos, M. Bonetti, D. Beysens, F. Perrot, 
T. Fr$\ddot{\rm o}$hlich, P. Carl$\grave{\rm e}$s, 
and B. Zappoli, Phys. Rev. E {\bf 57}, 5665 (1998). 




\bibitem{Beysens1} 
D. Beysens, Y. Garrabos, V. S. Nikolayev, 
C. Lecoutre-Chabot, J.-P. 
Delville,  and J. Hegseth, 
Europhys. Lett., {\bf 59},  245 (2002). 



\end{references}

\newpage 
\begin{figure}[tb]
\includegraphics[scale=0.8]{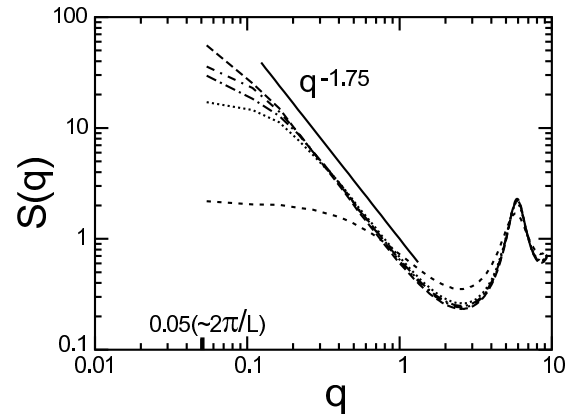}
\caption{The structure factor $S(q)$ 
at@ $n =0.37$ for 
 $T=0.65$ (short-dashed line at bottom), 
$0.51$, $0.5$, $0.495$, and $0.49$ (dashed line on top).
A line with a slope of $-7/4$ is included as a guide.
}
\label{oz}
\end{figure}

\begin{figure}[b]
\includegraphics[scale=0.8]{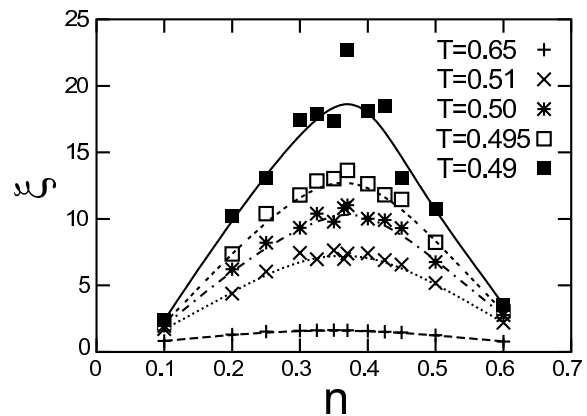}
\caption{
The correlation length $\xi$ vs the density 
at various temperatures  
obtained from the structure factor.
}
\label{xi}
\end{figure}

\begin{figure}[b]
\includegraphics[scale=0.8]{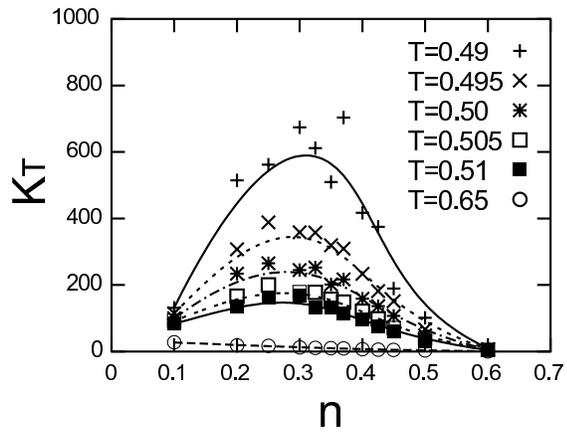}
\caption{The isothermal compressibility $K_T$ 
vs the density at various temperatures  
obtained from the structure factor.
}
\label{kt}
\end{figure}

\begin{figure}[tb]
\includegraphics[scale=0.6]{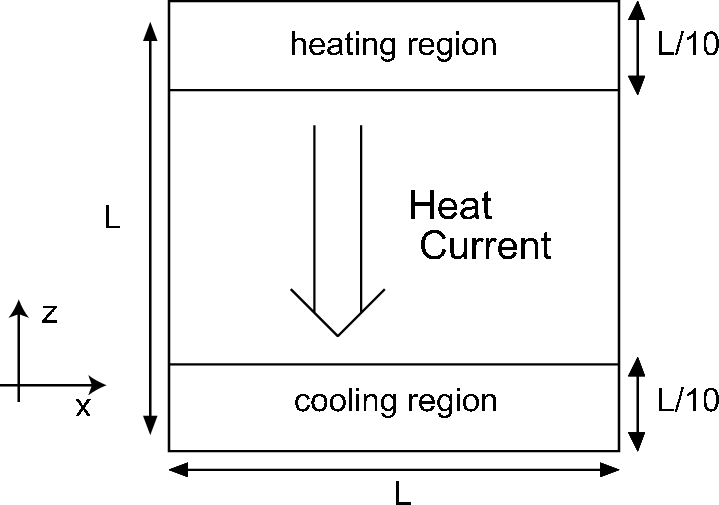}
\caption{Simulation cell under  heat flow 
composed of 
cooling, heating, and interior regions\cite{ohara}.}
\label{nemd}
\end{figure}

\begin{figure}[tb]
\includegraphics[scale=1]{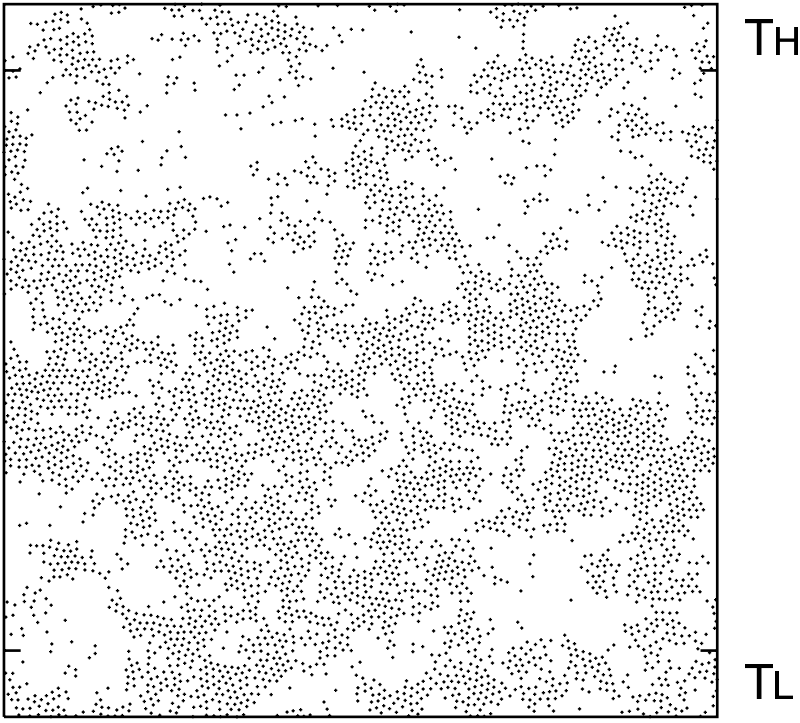}
\caption{Snapshot of the particle configuration 
 at $n=0.37$ in a steady state 
with  $T_{\rm L}=0.50$ and $T_{\rm H}=0.505$. 
The horizontal  bars at the vertical box lines 
mark the boundary  between the interior region and 
cooling or heating region. 
 }  
\label{nemd}
\end{figure}

\begin{figure}[tb]
\includegraphics[scale=0.8]{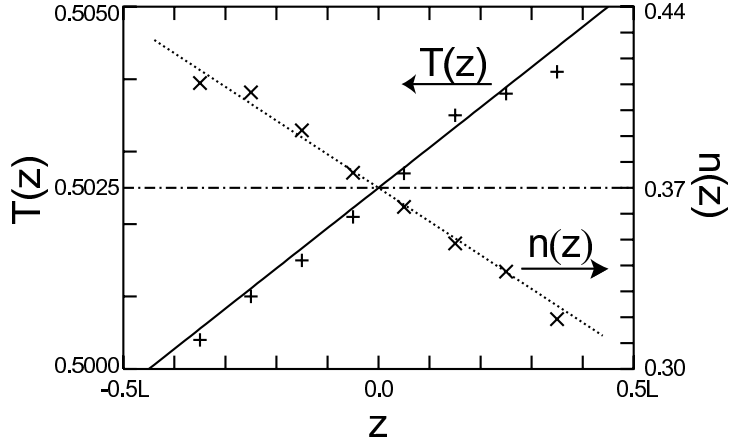}
\caption{
Steady-state temperature and density 
profiles in  the $z$-direction 
obtained by the time average. The solid  line   
 has  a slope of $\Delta T/(9L/10)$ with  $\Delta T=0.005$, 
 while the dotted line has a slope of 
$-n\alpha_p \Delta T/(9L/10)$ in Eq.16. 
}
\label{prof}
\end{figure}

\begin{figure}[b]
\includegraphics[scale=0.8]{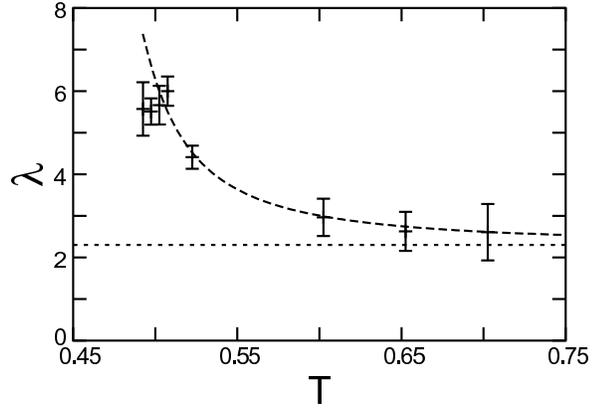}
\caption{The thermal conductivity $\lambda$ 
calculated from Eqs.16 and 17 
at $n=0.37$ for 
$T=0.7$, $0.65$, $0.6$, 
$0.52$, $0.51$, $0.5$, $0.495$, and $0.49$.
The bold dashed line is a view guide. The width of each  error bar 
is   twice of the variance of 10 data values corresponding to 
10 independent runs. 
}
\label{lambda}
\end{figure}

\begin{figure}[b]
 \includegraphics[scale=0.8]{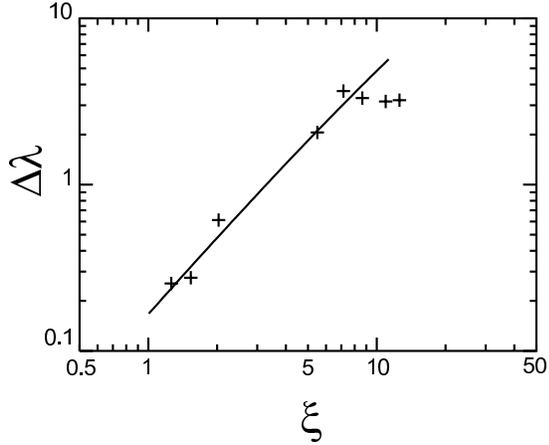}
 \caption{The singular part of the thermal conductivity 
$\Delta\lambda$ as a function of $\xi$
on logarithmic scales. 
The solid line is  the second term in Eq.19 with 
$A_\lambda=0.035$. 
}
\label{lambda-xi}
\end{figure}

\begin{figure}[tb]
\includegraphics[scale=0.8]{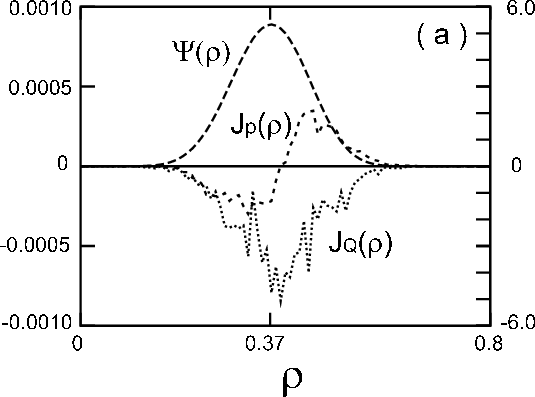}
\includegraphics[scale=0.8]{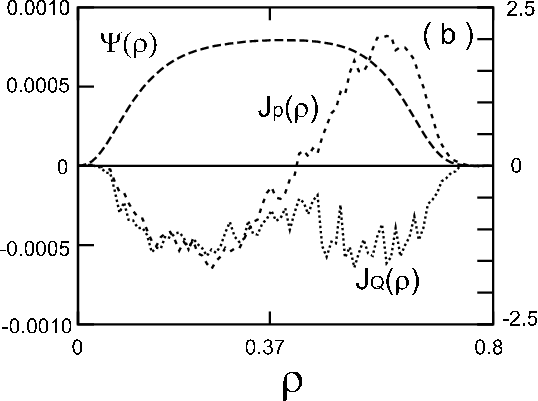}
\includegraphics[scale=0.8]{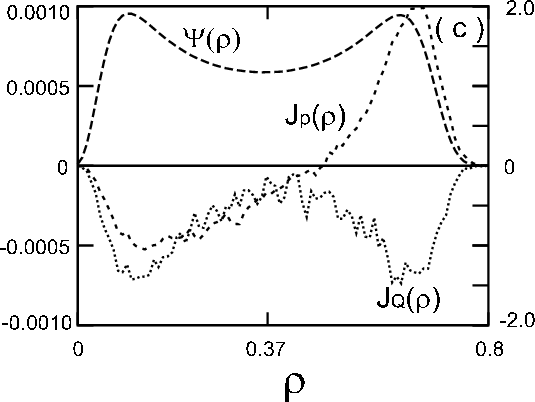}
\caption{(a) Density distribution function $\Psi(\rho)$
(right scale), 
momentum distribution $J_p(\rho)$, and heat-flux distribution 
 $J_Q(\rho)$ (left scale)
obtained at  $n=0.37$  for $T_{\rm L}=0.65$ in (a), 
 $T_{\rm L}=0.50$ in (b), and  $T_{\rm L}=0.48$ in (c).
}
\label{dist}
\end{figure}

\end{document}